# Remarks on the tunneling limit of strong-field photoionization

Jarosław H. Bauer *

Katedra Fizyki Teoretycznej Uniwersytetu Łódzkiego,
Ul. Pomorska 149/153, 90-236 Łódź, Poland

Some results from a recent work of Reiss [Phys. Rev. Lett. **101**, 043002 (2008); **101**, 159901(E) (2008).] are generalized for the case of elliptical polarization of a laser field. We also discuss the tunneling limit of strong-field photoionization.

---

*Electronic address: bauer@uni.lodz.pl

In his recent work [1] Reiss considered (among other things) physical significance of the so-called tunneling limit of ionization by lasers: $\gamma \to 0$ ($\gamma$ is the Keldysh adiabaticity parameter [2]). According to Reiss, the limit $\gamma \to 0$ applies *only* to ionization by quasistatic electric fields. In the present work we would like to state this more precisely, because in fact Ref. [1] concerns only *linear polarization* (LP). The latter one was mentioned never once in Reiss' work, hence it might suggest that the obtained results are valid for any polarization. (In fact, only the remark above Eq. (5) about a figure-8 pattern bespeaks LP in Ref. [1].) A freely strong plane-wave electromagnetic field is completely described by its frequency $\omega$, intensity $I$, and polarization. But only two ($\omega$, $I$) of these three quantities were evidently taken into account in Ref. [1]. The third important parameter, omitted in the analysis done by Reiss, is the laser field ellipticity. The main aim of the present work is to show that some conclusions from Ref. [1] (those regarding the limit $\gamma \to 0$) are valid only for *linear polarization* (LP). It appears that one can easily generalize them for any elliptical polarization. We also show that the case of *circular polarization* (CP) is qualitatively different from that of LP. In particular, one of the main conclusions from Ref. [1] about the onset of magnetic-field effects in the limit $\gamma \to 0$ does not apply for CP. In what follows we use atomic units: $\hbar = e = m_e = 1$ (substituting explicitly $-1$ for the electronic charge), and the same notation as in Ref. [1].

Ref. [1] is based on a classical relativistic dynamics describing a free point charge (an electron) moving in the monochromatic plane-wave laser field. On the other hand, the well-known early tunneling theories [2-8] assumed the non-relativistic and dipole (or long wavelength) approximation in the description of an outgoing electron. In the present work, unlike in Ref. [1], we assume that a tunneling theory should give the ionization or detachment rate in a laser field, when the outgoing electron tunnels through the Coulomb barrier (i.e. when tunneling is a dominant mechanism of ionization or detachment). In our opinion, a tunneling theory does not have to neglect the magnetic-field component of the laser [9], as it is assumed in Ref. [1]. There are numerous examples confirming our opinion in the recent review article of Popov [10]. Of course, for extremely strong fields relativistic treatment of tunneling [10-14] is necessary for any kind of polarization. In Fig. 1 of Ref. [12] there are specified three important regions of the laser field parameters ($\omega$, $I$), where the ionization or detachment can take place. In the intermediate range of these parameters one can keep the non-relativistic theory, but one has to take into account the magnetic-field component of the laser. Then the electric field $\vec{E}$ and the magnetic field $\vec{B}$ depend only on time and obey the condition $\vec{B} = \hat{n} \times \vec{E}$ ($\hat{n}$ is a unit vector in the propagation direction). In the present work we reconsider the area marked as "Magnetic Field Important" in Fig. 1 of Ref. [12].

Let us assume that the field propagates along the $x$-axis and its wave vector is given by $\vec{k} = k\hat{e}_x = [k,0,0]$ ($\hat{e}_x, \hat{e}_y, \hat{e}_z$ are real unit vectors; $k = \omega/c$; $\vec{r} = [x, y, z]$ in the laboratory frame). The field of an arbitrary polarization can be described by the following vector potential (and the scalar potential equal to zero)

$$\vec{A}(\vec{r},t) = a\left(\hat{e}_z \cos(\omega t - kx)\cos(\delta/2) \pm \hat{e}_y \sin(\omega t - kx)\sin(\delta/2)\right), \tag{1}$$

where $\delta$ is the ellipticity parameter ($\delta \in [0, \pi/2]$, and the signs $\pm$ correspond to two different helicities). The motion of the charge in the field given by Eq. (1) can be found exactly for any $\delta$. For LP ($\delta = 0$), and for CP ($\delta = \pi/2$) there are solutions to this problem in Sec. 48 (p. 134) of Ref. [15]. We generalize these solutions for any polarization here, and in the simplest frame of reference we get the following result

$$x = \frac{a^2}{8c\omega\varepsilon^2} \cos\delta \sin 2(\omega t - kx), \tag{2a}$$

$$y = \mp \frac{a}{\omega\varepsilon} \sin(\delta/2)\cos(\omega t - kx), \tag{2b}$$

$$z = \frac{a}{\omega\varepsilon} \cos(\delta/2)\sin(\omega t - kx), \tag{2c}$$

where $\varepsilon = \sqrt{c^2 + a^2/(2c^2)}$. (The simplest frame of reference is the one in which the charge is at rest on the average.) The amplitude of motion in the propagation direction, in Eq. (2a), may be treated as a direct measure of the effect of the magnetic field. In this way we generalize here Eq. (5) from Ref. [1] obtaining the following equation

$$\beta_0 = \frac{a^2}{8c\omega\varepsilon^2}\cos\delta = \frac{z}{2c(1+z_f)}\cos\delta \approx \frac{z}{2c}\cos\delta, \qquad (3)$$

where $z_f \equiv 2U_P/c^2 \equiv 2z\omega/c^2 \ll 1$ ($z$ and $z_f$ are the intensity parameters, and $U_P$ is the ponderomotive potential). In Eq. (3) we used expressions for the laser field intensity in atomic units ($I = (a\omega/c)^2 = 4z\omega^3$, which are the same for all $\delta$) and Eqs. (3) and (6) from Ref. [1]. Equations (2) and (3) are the main result of the present paper. For LP Eq. (3) reduces to Eq. (5) from Ref. [1], and for CP one obtains $\beta_0 = 0$ from Eq. (3). Therefore Eq. (8) from Ref. [1] is generalized here to

$$\beta_0 = 1 \quad \Rightarrow \quad I = 8c\omega^3/\cos\delta, \qquad (4)$$

for $0 \leq \delta < \pi/2$. Figure 1 (which is similar to Fig. 2 from Ref. [1]) shows the line $z_f = 1$ and the lines $\beta_0 = 1$ from Eq. (3) for different ellipticities. For $\delta \neq 0$ the line $\beta_0 = 1$ moves towards the line $z_f = 1$ with increasing $\delta$. However, due to logarithmic scales in Fig. 1, for typical ellipticities different from LP and CP ($0 < \delta < \pi/2$), the area where magnetic-field effects become important diminishes only very slightly with increasing $\delta$. For CP ($\delta = \pi/2$) the intermediate area of the parameters $\omega$, $I$ disappears. This means that one should use qualitatively different description (the fully relativistic one) of strong-field ionization beginning in the vicinity of the line $z_f = 1$ for CP. The plane-wave laser field is a transverse field. For CP the magnetic-field component of the Lorentz force acting on the charge is equal to zero because this component is always parallel to the velocity of the charge. Although the magnetic field in the circularly polarized plane-wave may be freely strong, it can not force the charge to move along the propagation direction. For CP, even in the fully relativistic regime, the motion takes place along a circle lying in the polarization plane. Equations (2) show that for $0 < \delta < \pi/2$ the relativistic charge moves in three spatial dimensions (in the simplest frame of reference), but for $\delta = 0$ or $\delta = \pi/2$ - only in two dimensions. For example, for CP one obtains from Eqs. (2) that for $\omega = const$ and $I \to \infty$ the radius of the above mentioned circle approaches a finite limit $c/\omega$ instead of infinity predicted by the non-relativistic dynamics. Therefore, for any $\delta$, the limit $\gamma \to 0$ for extremely strong laser fields requires relativistic treatment.

By the way, let us note that the non-relativistic limit $\gamma \to 0$, when only the electric field vector (in the dipole approximation) is present, may have a very well-defined physical meaning. For example, the effect of a slowly rotating (with the frequency $\omega$) electric field vector (of a constant length) on the bound system in the limit $\omega \to 0$ should be the same as for the static electric field. Therefore the ionization rates for CP and $I = const$, calculated in the non-relativistic and dipole approximations (hence $\vec{E} = \vec{E}(t)$ and $\vec{B} = \vec{0}$) should coincide with the static field ionization rates. By analogy, the ionization rates for LP and $I = const$, in the same approximations, should coincide with the static field ionization rates averaged over a field period [16]. We have recently studied numerically the $S$-matrix theory of strong-field photoionization in this context for the $H(1s)$ atom and both polarizations [17,18]. Furthermore, recently Vanne and Saenz [19] have shown analytically that in the so-called velocity gauge in the low-frequency limit the ionization rate for LP is proportional to the laser frequency. As a result, $I = const$ and $\omega \to 0$ lead to nulling of the ionization rate. In our opinion, this clearly *unphysical* result is a consequence of the lack of gauge invariance of the ionization probability amplitude in this case [20]. In the limit $I = const$ and $\omega \to 0$ (then also $\gamma \to 0$) the ratio $U_P/E_B$ goes to infinity. The latter fact should be the sufficient applicability condition for the non-relativistic $S$-matrix theory which assumes that $\vec{E} = \vec{E}(t)$ and $\vec{B} = \vec{0}$. When the approximate, but gauge-invariant Keldysh's theory [2] is used, one obtains at least an order of magnitude (or better) agreement with the exact static field results of Scrinzi [21] for the $H(1s)$ atom. As one should expect, the agreement usually improves with increasing the ratio $U_P/E_B$ (see Fig. 5 in Ref. [17] and Figs. 1, 2 in Ref. [20]).

As a side remark, we note that even for LP the magnetic field effects do not have to be strong in the intermediate regime of the parameters $\omega$ and $I$ (the area between the line $z_f = 1$ and the line $\beta_0 = 1$ and $\delta = 0$ in Fig. 1). In a recent experiment [22] with argon and pulsed-laser 800 $nm$ radiation at an intensity of up to $10^{19}$ $W/cm^2$ (cp. also Fig. 2 from Ref. [1]) the average Keldysh parameter $\gamma$ was equal to $0.03$. The authors of Ref. [22] confirmed validity of the well-known ADK/WKB tunneling model [3-8] (which is the non-relativistic and dipole approximation theory) for the above parameters of the strong laser field. According to authors of Ref. [22], the results of their experiment may be interpreted within a two-step model, where the initial tunneling ionization process is dominated by the non-relativistic effects while the photoelectron continuum dynamics are strongly relativistic.

## ACKNOWLEDGMENTS

The author is indebted to Professor Howard R. Reiss for his correspondence and for arousing author's interest in the subjects related to this work. The present paper has been supported by the University of Łódź.

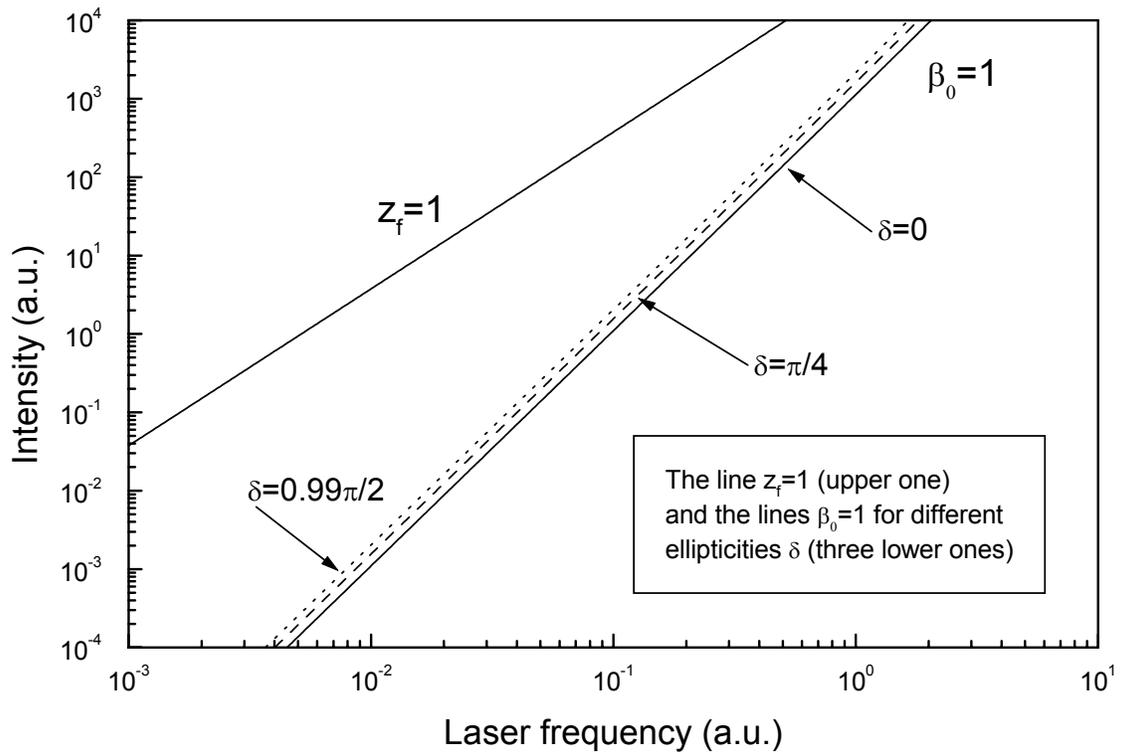

FIG. 1. The slanted lines indicate the intensity as a function of frequency, as a result of fixing either of two parameters. The line $z_f = 1$ corresponds to any ellipticity parameter $\delta$. The lines $\beta_0 = 1$ correspond to $\delta = 0$ (linear polarization), $\delta = \pi/4$ (elliptical polarization), and $\delta = 0.99\pi/2$ (almost circular polarization). (See the text for more details.)